\documentclass[12pt]{article}
\usepackage{amsmath,amssymb} 
\usepackage{subcaption}   
\DeclareMathOperator{\arccoth}{arccoth}
\DeclareMathOperator{\arctanh}{arctanh}
\DeclareMathOperator{\arcsinh}{arcsinh}
\usepackage{amsfonts}
\usepackage{enumerate}
\usepackage{ytableau}
\usepackage{hyperref}
\usepackage[
backend=biber,
style=numeric-comp,
sorting=none
]{biblatex}
\usepackage{bm}
\usepackage{tikz}
\usepackage{blkarray}  
\begin{document}

\def\im{\text{i}}
\def\eqa{\begin{eqnarray}}
\def\eqae{\end{eqnarray}}
\def\be{\begin{equation}}
\def\ee{\end{equation}}
\def\bea{\begin{eqnarray}}
\def\eea{\end{eqnarray}}
\def\ba{\begin{array}}
\def\ea{\end{array}}
\def\bd{\begin{displaymath}}
\def\ed{\end{displaymath}}
\def\eg{{\it e.g.~}}
\def\ie{{\it i.e.~}}
\def\Tr{{\rm Tr}}
\def\tr{{\rm tr}}
\def\>{\rangle}
\def\<{\langle}
\def\a{\alpha}
\def\b{\beta}
\def\c{\chi}
\def\del{\delta}
\def\e{\epsilon}
\def\f{\phi}
\def\vf{\varphi}
\def\tvf{\tilde{\varphi}}
\def\g{\gamma}
\def\h{\eta}
\def\j{\psi}
\def\k{\kappa}
\def\l{\lambda}
\def\m{\mu}
\def\n{\nu}
\def\w{\omega}
\def\p{\pi}
\def\q{\theta}
\def\r{\rho}
\def\s{\sigma}
\def\t{\tau}
\def\u{\upsilon}
\def\x{\xi}
\def\z{\zeta}
\def\D{\Delta}
\def\F{\Phi}
\def\G{\Gamma}
\def\J{\Psi}
\def\L{\Lambda}
\def\W{\Omega}
\def\P{\Pi}
\def\Q{\Theta}
\def\S{\Sigma}
\def\U{\Upsilon}
\def\X{\Xi}
\def\nab{\nabla}
\def\pa{\partial}
\def\Renyi{R$\acute{\text{e}}$nyi }
\def\Poincare{Poincar$\acute{\text{e}}$ }
\def\Banados{Ba$\tilde{\text{n}}$ados }
\newcommand{\lra}{\leftrightarrow}

\newcommand{\bc}{{\mathbb{C}}}
\newcommand{\br}{{\mathbb{R}}}
\newcommand{\bz}{{\mathbb{Z}}}
\newcommand{\bp}{{\mathbb{P}}}

\def\({\left(}
\def\){\right)}
\def\[{\left[}
\def\]{\right]}
\def\nn{\nonumber \\}
\def\nt{\notag \\}
\def\bal#1\eal{\begin{align}#1\end{align}}

\newcommand{\red}{\textcolor[RGB]{255,0,0}}
\newcommand{\blue}{\textcolor[RGB]{0,0,255}}
\newcommand{\green}{\textcolor[RGB]{0,255,0}}
\newcommand{\cyan}{\textcolor[RGB]{0,255,255}}
\newcommand{\magenta}{\textcolor[RGB]{255,0,255}}
\newcommand{\yellow}{\textcolor[RGB]{255,255,0}}
\newcommand{\sky}{\textcolor[RGB]{135, 206, 235}}
\newcommand{\orange}{\textcolor[RGB]{255, 127, 0}}
\def\d{\operatorname{d}}
\def\ttbar{T$\overline{\text{T}}$ }

\title{\textbf{Aspects of three-dimensional C-metric}}
\vspace{14mm}
\author{Jia Tian$^{1,2}$\footnote{wukongjiaozi@ucas.ac.cn} and Tengzhou Lai$^{3}$\footnote{laitengzhou20@mails.ucas.ac.cn}}
\date{}
\maketitle

\begin{center}
	{\it $^1$State Key Laboratory of Quantum Optics and Quantum Optics Devices, Institute of Theoretical Physics, Shanxi University, Taiyuan 030006, P.~R.~China\\
 \vspace{2mm}
		$^2$Kavli Institute for Theoretical Sciences (KITS),\\
		University of Chinese Academy of Science, 100190 Beijing, P.~R.~China \\
  \vspace{2mm}
  $^3$School of Physical Sciences, University of Chinese Academy of Sciences, Zhongguancun East Road 80, Beijing 100190, P.~R.~China
	}
\vspace{10mm}
\end{center}

\date{}
\maketitle

\begin{abstract}
In this work, we present an extensive analysis of the thermodynamics and holographic properties of three-dimensional C-metrics in the FG gauge, where we find that the free energy is equal to the Euclidean on-shell action with a generic conformal factor. For the black hole solutions we find that  Smarr relation and the first law of thermodynamics can be formulated when the contributions of the boundary entropy are considered . We also compute holographic entanglement entropy following the AdS/BCFT formalism. By comparing the free energies of different bulk solutions with a fixed flat torus boundary geometry, we find that a specific type of accelerating black hole is dominant in the high temperature regime.
\end{abstract}

\baselineskip 18pt
\newpage

\tableofcontents
\section{Introduction}
The four-dimensional (4D) C-metric is a class of exact solutions of Einstein equations which has been known for many years \cite{levi,weyl}. It describes black holes with acceleration due to the pulling of the cosmic strings that are represented by conical singularities \cite{Kinnersley:1970zw,bonnor}\footnote{More details about the interpretation to this kind of solutions can be found for example in \cite{Griffiths:2006tk}}. Although the solutions are not smooth,  nonetheless the black hole+cosmic brane systems satisfy usual laws of black hole thermodynamics \cite{Appels:2016uha}. As a black hole model, the 4D C-metric  have been investigated in many aspects \cite{Letelier:1998rx,Bicak:1999sa,Podolsky:2000at,Pravda:2000zm,Dias:2002mi,Griffiths:2005qp,Krtous:2005ej,Dowker:1993bt,Emparan:1999wa,Emparan:1999fd,Emparan:2000fn,Gregory:2008br,Emparan:2020znc,Lu:2014sza,Ferrero:2020twa,Cassani:2021dwa,Ferrero:2021ovq,Astorino:2016xiy,Appels:2017xoe,Anabalon:2018ydc,Anabalon:2018qfv,EslamPanah:2019szt,Gregory:2019dtq,Ball:2020vzo,Ball:2021xwt,Gregory:2020mmi,Kim:2023ncn,Clement:2023xvq,Hubeny:2009kz,Ferrero:2020laf,Ferrero:2021etw,Boido:2022iye,Ashtekar:1981ar}. Despite these progress, the quantum properties and especially its holographic correspondence are still relatively little known. Recently, by a direct dimension truncation of the 4D C-metric, (non-rotating) 3D C-metrics are constructed in \cite{Arenas-Henriquez:2023hur} \footnote{For early studies of 3D accelerating black holes, see also \cite{Xu:2011vp,Astorino:2011mw} and for 3D C-metrics not in the pure Einstein gravity see \cite{EslamPanah:2023ypz,EslamPanah:2023rqw}.}, which take the general form:
\bea 
ds^2&=&\frac{1}{\Omega^2}\left[-P(y)a^2dt^2+\frac{dy^2}{P(y)}+\frac{dx^2}{Q(x)}\right],\\
\Omega &=& a(x-y),
\eea 
and come with three classes:
\be \label{eq:3class}
\begin{tabular}{|c|c|c|c|}
\hline
Class  & $Q(x)$ & $P(y)$ & Maximal range of $x$ \\
\hline
I & $1-x^2$ & $\frac{1}{a^2}+(y^2-1)$ & $|x|<1 $\\
II & $x^2-1$ & $\frac{1}{a^2}+(1-y^2)$ & $x<-1 $ or $x>1$\\
III & $1+x^2$ & $\frac{1}{a^2}-(1+y^2)$ & $\mathbb{R} $\\
\hline
\end{tabular}
\ee 
Here, the parameter $a$ characterizes the acceleration and the AdS radius $\ell$ has been set to $\ell=1$. The 3D C-metrics are expected to serve as a simplified research framework for exploring the quantum or holographic features of acceleration. As nicely demonstrated in \cite{Arenas-Henriquez:2023hur}, each class  describes a particular patch of the global AdS$_3$ spacetime. The boundaries of the patch can either be a horizon, the conformal boundary or a EOW brane. The locations of the horizon are determined by the zeros of $P(y)$. The conformal boundary is located at $x-y=\epsilon(x)$, where $\epsilon(x)$ is a UV cut-off. In the C-metric, the profile of the EOW is extremely simple which is just a constant-$x$ surface and the tension of the brane is related to the position of the brane via\footnote{We used the standard convention in AdS/BCFT\cite{BCFT1,BCFT2}, the brane tension here is $4\pi G_N $ times the one defined in  \cite{Arenas-Henriquez:2023hur}.} $\mu=\pm a \sqrt{1-x^2}$. Obviously, the maximal C-metrics \eqref{eq:3class} are all bounded by the tensionless EOW branes at $x=\pm 1$. One can also consider non-maximal C-metrics and by gluing them along the EOW branes more interesting geometries can be constructed as shown in \cite{Arenas-Henriquez:2023hur} which can describe accelerating particles or accelerating black holes. It turns out that if we fix the asymptotic 2D Euclidean boundary to be a torus, there are at least seven possible bulk solutions.
The geometry aspects of these solutions have been carefully analyzed. However, the thermodynamical and holographic properties have not been fully understood. One of the complications is that AdS$_3$ gravity suffers from a holographic Weyl anomaly \cite{Henningson:1998gx} which can be reflected in the ambiguity of the FG gauge \cite{graham1985charles,Fefferman:2007rka} of C-metrics. As a result, holographic mass and thermodynamical first law depends on the conformal representative of the boundary metric. To make progress, the authors in \cite{Arenas-Henriquez:2022www} concede to focus on a special choice of the gauge as known as the ADM gauge \cite{arnowitt1961coordinate} in which they show the Euclidean on-shell action equates to the free energy but the first law is not able to be derived. In \cite{Tian:2023}, we point out that EOW brane will also contribute to the black hole entropy and by treating the black hole+EOW brane as a whole system, a first law is possible to be formulated. 

In this work, we show how to use FG gauge to systematically study the holographic properties of the 3D C-metric in which all the conformal representatives are put on the same footing. We will focus on the example of the class I solution since the analysis on other cases are parallel and some of details about other solutions will be presented in the appendix. The paper is organized as follows: In section 2, we derive the holographic mass in the FG gauge and show it agrees with the Euclidean on-shell action calculation for all the choice of conformal gauges. We find the maximal class I solution in the slow phase can be understood as a one-parameter interpolation between the global AdS and the \Poincare AdS space. We also show the spherical coordinates of the C-metric may not have a well-defined boundary description. In section 3, we study the holographic entanglement entropy in a C-metric wedge within the formalism of AdS/BCFT duality. In section 4, we fix the boundary to be torus and then study the possible phase transitions of the bulk solutions by comparing their free energies. 

\section{Class I: accelerating particle}
For a maximal class I patch, when $0<a<1$ (in the slow phase), $P(y)$ has no real roots so there is no Killing horizon in the patch. To better view the geometry, it is convenient to introduce the spherical coordinates
\be 
r=-(ay)^{-1},\quad \phi=\arccos(x),\quad |x|\leq 1,
\ee 
such that the metric can be written as
\bea  \label{eq:class1_smetric}
ds^2 &=& \frac{1}{(1+a r\cos\phi)^2}\left[-f(r){dt^2}{}+\frac{dr^2}{f(r)}+r^2d\phi^2\right], \\
f(r) &=& 1+(1-a^2)r^2,
\eea
which reduces to the standard global AdS coordinates in the limit $a=0$. However, the angular coordinate $\phi$ has a range of $[0,\pi)$ so the geometry only covers ``half " of the AdS space just like a Rindler wedge only covers ``half " of the space. Then it is natural to glue two copies of such solution along the EOW brane at $x=1$ ($\phi=0$) and at $x=-1$ ($\phi=\pi $). As we mentioned above, the constant-$x$ surface describes a EOW brane with tension\footnote{where the $\pm$ depends on the direction of the normal vector.}
\be \label{eq:tension}
\mu=\pm a\sqrt{1-x^2}.
\ee  

\begin{figure}[ht]
\centering 
  \includegraphics[scale=1.5]{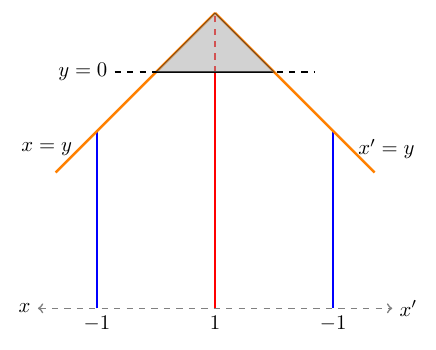}
  \caption{Geometry of Class I solution with no Killing horizon. The horizontal and vertical direction represents $x$ and $y$ axis respectively and the region filled with gray color is not covered by spherical coordinates since $y=0$ corresponds to $r=\infty$.}\label{fig:geometry}
\end{figure}

To understand the result, we can refer to the coordinate transformation between \eqref{eq:class1_smetric} and the global coordinates.  All AdS$_3$ solutions can be embedded in to a hyperboloid $\mathbb{R}^{2,2}$ defined by
\be
-X_0^2+X_1^2+X_2^2-X_3^2=-1.
\ee 
For example, the global AdS$_3$ is embedded as
\bea 
&&X_0=\sqrt{1+R^2}\sin{T}{},\quad X_1=R\sin\Theta,\\
&&X_3=\sqrt{1+{R^2}{}}\cos{T}{},\quad X_2=R\cos\Theta,
\eea 
and the metric in global AdS$_3$ is
\bea \label{eq:global}
ds^2=-(1+{R^2}{})dT^2+\frac{dR^2}{(1+{R^2}{})}+R^2d\Theta^2.
\eea 
The class I solution in the slow phase can be embedded as \cite{Arenas-Henriquez:2022www}
\bea 
&& X_0=\frac{a\sqrt{P}}{\sqrt{1-a^2}\Omega}\sin \sqrt{1-a^2}t,\quad X_1=\frac{\sqrt{Q}}{\Omega},\\
&& X_3=\frac{a\sqrt{P}}{\sqrt{1-a^2}\Omega}\cos \sqrt{1-a^2} t,\quad X_2=\frac{1}{\Omega}(\sqrt{1-a^2}x+\frac{a^2y}{\sqrt{1-a^2}}),
\eea 
so by comparing these embeddings the coordinate transformation between the global AdS$_3$ and the C-metric can be obtained as follows
\bea \label{eq:globalToC}
T=\sqrt{1-a^2}t,\,\, R=\frac{\sqrt{1-x^2+(\sqrt{1-a^2}x+\frac{a^2 y}{\sqrt{1-a^2}})^2}}{a(x-y)},\,\,\tan\Theta=\frac{\sqrt{1-x^2}}{\sqrt{1-a^2}x+\frac{a^2 y}{\sqrt{1-a^2}}}.
\eea 
From these transformation it is easy to find that the constant-$x$ surface in the C-metric  corresponds to the profile
\be \label{eq:eow}
R=\frac{a\sqrt{1-x^2}}{\sqrt{1-a^2}\sqrt{1-x^2}\cos\Theta-x\sin\Theta}=-\frac{a\sqrt{1-x^2}}{\sqrt{1-a^2+a^2x^2}}\frac{1}{\sin(\Theta-\Theta_*)},
\ee 
in the global coordinates, where the shifted angle $\Theta_*$ is determined by
\be 
\cos\Theta_*=\frac{x}{\sqrt{1-a^2+a^2x^2}}.
\ee 
The profile \label{eq:eow} exactly describes the EOW  in global AdS with tension \eqref{eq:tension}\footnote{Recall that a EOW with tension $\mu$ has a profile: $R \sin\Theta=\pm \frac{\mu}{\sqrt{1-\mu^2}}$.}.
We also find that the center point $r=-(ay)^{-1}=0$ in the C-metric is mapped to a point at
\be \label{eq:map0}
R=\frac{a}{\sqrt{1-a^2}},\quad \Theta=-\pi
\ee 
in the global AdS. This implies the observer sitting in the center of the C-metric has to be accelerated to counter with the negative pressure of the AdS space. The boundary of the spherical coordinates $r=\infty$  also corresponds to a EOW brane with tension $\mu_\infty=\sqrt{1-a^2}$ in the global coordinate
\bea 
R=\frac{\sqrt{1-a^2}}{a\cos\Theta}
\eea 
which passes through the antipodal point \be 
R=\frac{a }{\sqrt{1-a^2}},\quad \Theta=0.
\ee
The induced metric on this EOW brane is
\be 
ds^2_{r\rightarrow \infty}=\frac{1}{a^2\cos^2\phi}\(-(1-a^2)dt^2+d\phi^2\).
\ee 
 Therefore, when $\phi\in (-\frac{\pi}{2},\frac{\pi}{2})$, the spherical coordinates can not reach the conformal boundary as shown in Fig.\eqref{fig:geometry}. By computing the on-shell action, we will show that the  patch described by the spherical coordinates may not have a boundary dual.

\subsection{Holographic mass: FG expansion}
First we want to derive the holographic mass by rewriting the metric into the FG gauge.
As shown in \cite{Anabalon:2018ydc,Arenas-Henriquez:2022www}, to solve the FG gauge  we can take the following ansatz
\be
y= \xi+\sum_{m=1} F_m(\xi )z^m ,\quad 
x= \xi+\sum_{m=1} G_m(\xi )z^m,
\ee
and determine the unknown functions  by comparing with FG gauge order by order. The result is 
\be 
ds^2=\frac{dz^2}{z^2}+\frac{g^{(0)}_{ij}dx^idx^j}{z^2}+ g^{(2)}_{ij}dx^idx^j+z^2 g^{(4)}_{ij}dx^idx^j
\ee 
with
\bea 
g^{(0)}_{ij}dx^idx^j&=&\omega(\xi)^2\left(-dt^2+\frac{d\xi^2}{(1-\xi^2)\Gamma(\xi )^2}\right),\\
\omega^2&=&\frac{\Gamma^3}{F_1^2 a^2},\quad \Gamma=1-a^2+a^2\xi^2, \label{eq:gamma}
\eea 
and 
\bea  
&&g_{tt}^{(2)}=-\frac{1-a^2}{2}-\frac{(1-\xi^2)\Gamma^2\omega'^2}{2\omega^2},\\
&&g_{\xi \xi }^{(2)}=-\frac{(1-a^2)}{2(1-\xi^2)\Gamma^2}-\frac{(1-3a^2(1-\xi^2))\xi \omega'}{\Gamma(1-\xi^2)\omega}+\frac{(2\omega\omega''-3{\omega'}^2)}{2\omega^2},\\
&&g^{(4)}=\frac{1}{4}g^{(2)}{g^{(0)}}^{-1}g^{(2)}.
\eea 
To have a real conformal factor, \eqref{eq:gamma} implies that
\be 
\Gamma>0,\quad \rightarrow \quad \xi^2>\frac{a^2-1}{a^2},
\ee 
which is always satisfied in the slow phase while it will impose addition constraint in the rapid phase ($a>1$).
For a non-trivial conformal factor, the boundary metric $g^{(0)}$ is not flat. The Ricci scalar seems to be very complicated:
\be
R[g^{(0)}]=\frac{2\Gamma}{\omega^4}\((3\Gamma-2)\xi \omega \omega'+\Gamma(1-\xi^2)(\omega'^2-\omega \omega'')\),
\ee 
but it is well-known that in 2D the combination $\sqrt{g^{(0)}}R[g^{(0)}]$ is a total derivative:
\be \label{eq:total}
\sqrt{g^{(0)}}\,R[g^{(0)}]=\frac{d}{d\x}\(-\frac{2\G\sqrt{1-\x^2}\w'}{a \w}\).
\ee 
Below we will use this fact to simply some of the calculations. In the FG gauge, the holographic stress-energy tensor can be immediately derived from the relation\cite{deHaro:2000vlm,Skenderis:2002wp}
\bea 
\langle T_{\alpha \beta}\rangle=\frac{1}{8\pi G_N}(g^{(2)}_{\alpha \beta}-\text{Tr}[g^{(2)}]g^{(0)}_{\alpha \beta}),
\eea 
with the result
\bea 
&&8\pi G_N T_{\xi\xi }=-\frac{1-a^2}{2(1-\xi^2)\Gamma^2}-\frac{\omega'^2}{2\omega^2},\\
&&8\pi G_N  T_{tt}=\frac{1}{2}\left((a^2-1)+2(2\Gamma-3\Gamma^2)\xi \frac{\omega'}{\omega}+\Gamma^2(1-\xi^2)\frac{(2\omega \omega''-3\omega'^2)}{\omega^2} \right).
\eea 
The trace of stress-energy tensor gives the correct conformal anomaly \cite{bonora1986weyl,Deser:1993yx,Duff:1993wm}:
\be
\text{Tr}[T]=T_{\xi}^{\xi }+T_{\tau}^\tau
=\frac{1}{16\pi G_N}R[g^{(0)}].
\ee 
Using the relation and the identity \eqref{eq:total}, the holographic mass can be expressed in the following clean form
\bea 
M[\omega]&\overset{}=& -\int d\xi\sqrt{-g^{(0)}}T^\tau_\tau 
\\&=&-\frac{1}{16\pi G_N}\int_{}^{}d\xi \sqrt{-g^{(0)}}R[g^{(0)}] \label{eq:total2}\\
&& -\frac{1}{16\pi G_N}\int d\xi \(\frac{1-a^2}{\sqrt{1-\xi^2}\Gamma}+\frac{\sqrt{1-\xi^2}\Gamma \omega'^2}{\omega^2}{}\). \label{eq:holomass}
\eea 
When $\omega(\xi)=w$ is a constant so that $R[g^{(0)}]=0$,  we find 
\bea 
 M[\omega=w]=-\frac{1 }{16\pi G_N} \int d\xi \frac{1-a^2}{\sqrt{1-\xi^2}\Gamma}=-\frac{\sqrt{1-a^2}}{8G_N}.
\eea
In the zero acceleration limit $a\rightarrow 0$, it reduces to the mass of the global AdS solution
\bea 
M_{\text{global}}=-\frac{\sqrt{1-a^2}}{8G_N} \rightarrow|_{a=0}=- \frac{1}{8G_N}=-\frac{c}{12}.
\eea 
The technical subtlety of the integral in glued geometry needs to be emphasized here. When we evaluate the integral over $d\xi $, we always first change the variable 
\be 
\xi =\cos\phi,
\ee 
in the regime $\xi\in [-1,1],\phi\in[-\pi,0]$ so that
\be 
\int \frac{d\xi }{\sqrt{1-\xi^2}}=d\phi.
\ee 
In the end we continue the integration domain to $[-\pi,\pi]$. Since the angular variable $\phi$ is now periodic, we can disregard the total derivative term \eqref{eq:total2}. As a result, the expression for the holographic mass \eqref{eq:holomass} isolates the conformal representative dependent part, which solely originates from the conformal anomaly.
One can also rewrite the metric in the standard form the FG gauge:
\bea 
&&ds^2=\frac{dz^2}{z^2}+\frac{1}{z^2}e^\Phi dv d\bar{v}+\frac{1}{2}\mathcal{T}_\Phi dv^2+\frac{1}{2}\bar{\mathcal{T}}_\Phi d\bar{v}^2+\frac{1}{4}R_\Phi dv d\bar{v} \nonumber \\
&&\quad +\frac{1}{4}z^2 e^{-\Phi}(\mathcal{T}_\Phi dv+\frac{1}{4}R_\Phi d\bar{v})(\bar{\mathcal{T}}_\Phi d\bar{v}+\frac{1}{4}R_\Phi d{v}), \label{eq:standardFG}
\eea 
where
\bea 
\mathcal{T}_\Phi=\partial_v^2\Phi-\frac{1}{2}(\partial_v \Phi)^2+\mathcal{L}(v),\quad \bar{\mathcal{T}}_\Phi=\bar{\partial}_v^2\Phi-\frac{1}{2}(\bar{\partial}_v \Phi)^2+\bar{\mathcal{L}}(\bar{v}),\quad R_\phi=4\partial_v \bar{\partial}_v \Phi,
\eea 
and 
\bea 
&&v=X+t,\quad \bar{v}=X-t, \quad X=\frac{\text{arc}\tan[\frac{\xi }{\sqrt{1-a^2}\sqrt{1-\xi^2}}]}{\sqrt{1-a^2}}\in \frac{1}{\sqrt{1-a^2}}[-\frac{\pi}{2},\frac{\pi}{2}],\\
&&\Phi=2\log\omega,\quad \mathcal{L}=\bar{\mathcal{L}}=-\frac{1-a^2}{2}.
\eea 
When the boundary metric is flat, in the limit $a\rightarrow 0$ the geometry reduces to global AdS while in the limit $a\rightarrow 1$, the geometry approaches to \Poincare AdS. So we can interpret the geometry \eqref{eq:class1_smetric} as a one parameter interpolation between global AdS and \Poincare AdS. 
\subsection{On-shell action}
\label{sec:on_shell}
In this section we show that the extended bulk region has a well-defined on-shell action which agrees with the holographic mass while the spherical coordinates patch does not. The regularized Euclidean on-shell action has a bulk term
\bea 
I_{\text{bulk}}&=&-\frac{1}{16\pi G_N}\int d^3 x \sqrt{g}\(R+2\)\nonumber \\
&=&
\frac{1}{8\pi G_N a^2}\int dt_E dx\frac{1}{\sqrt{1-x^2}}\frac{1}{\epsilon^2 \Lambda(x)^2},
\eea 
where we have introduced the cut-off surface defined by 
\be \label{eq:cutoff}
y=x-\epsilon \Lambda(x),
\ee 
 and assumed that the range of $y$ variable is: $y\in(-\infty,x)$. This choice of UV cut off is not standard but it is more natural as we argued in the Appendix \eqref{appendix:cut-off} because it can correctly capture the ambiguity of the FG gauge. 
The boundary contribution of the Euclidean action is
\bea
I_{\text{bdy}}&=&-\frac{1}{8\pi G_N}\int dt_E dx\sqrt{h}(K-1)\\
&=&-\frac{1}{8\pi G_N a^2}\int dt_E dx \(\frac{1}{\sqrt{1-x^2}\epsilon^2 \Lambda(x)^2}+\frac{d (\frac{\sqrt{1-x^2}}{\Lambda(x)})}{dx}\frac{1}{\epsilon}+U(x)\)
\eea 
where $U(x)$ is very complicated but we can subtract a total derivative term to simplify it. The simplified result is
\bea  
U(x)+\frac{d}{dx}\(\frac{\sqrt{1-x^2}\Gamma\omega'}{\omega}\)&=&\frac{1-a^2}{2\sqrt{1-x^2}\Gamma}+\sqrt{1-x^2}\Gamma\frac{\omega'}{2\omega^2}\nonumber \\
&=&\sqrt{h}\(\frac{a^2(1-a^2)}{2\ell \omega^2}+\frac{a^2(1-x^2)\Gamma^2\omega'^2}{2\ell \omega^4}\)\equiv U_0, 
\eea  
where
\be 
\omega=\frac{\sqrt{1-a^2+a^2 x^2}}{\Lambda(x)}=\frac{\sqrt{\Gamma}}{\Lambda(x)}.
\ee 
Adding two terms together gives the total on-shell action
\be 
I_{\text{total}}=-\frac{1}{16\pi G_N}\int d^2x\sqrt{h}\(\frac{1-a^2}{\omega^2}+\frac{(1-x^2)\Gamma^2\omega'^2}{\omega^4}\)=\int dt_E M[\omega],
\ee 
as expected. Note that the $1/\epsilon$ does not contribute because the range of $x$ is $x\in [-1,1]$. Otherwise, we should introduce another term
\bea \label{eq:onshelleow}
\frac{1}{4\pi G_N a^2}\int dt_E\(\frac{\sqrt{1-x^2}\omega}{\sqrt{\Gamma}}\)\Big|^{x_f}_{x_i}.
\eea 

For the spherical coordinate patch, we need to subtract a region as shown in Fig.\eqref{fig:geometry}.  Presumably we should add a EOW at $y=0$.  The on-shell action in this region is
\bea  
I_{\text{sub}}&=&-\frac{1}{8\pi G_N a}\int dt_E dx\, U_0(x)+\frac{1}{8\pi G_N a}\int dt_E dx\frac{\sqrt{1-a^2}-1}{x^2\sqrt{1-x^2}} \nonumber \\
&+&\frac{1}{8\pi G_N a}\int dt_E\(\frac{1}{\epsilon}\frac{\omega(0)}{\sqrt{1-a^2}}-\frac{(1-a^2)\omega'(0)}{\omega(0)}\),
\eea   
which is divergent no matter with or without the EOW brane action. Therefore, we conclude the patch described by the spherical coordinates \eqref{eq:class1_smetric} does not have a simple boundary dual.

\subsection{Conical geometry}
\begin{figure}[ht]
\centering 
  \includegraphics[scale=1]{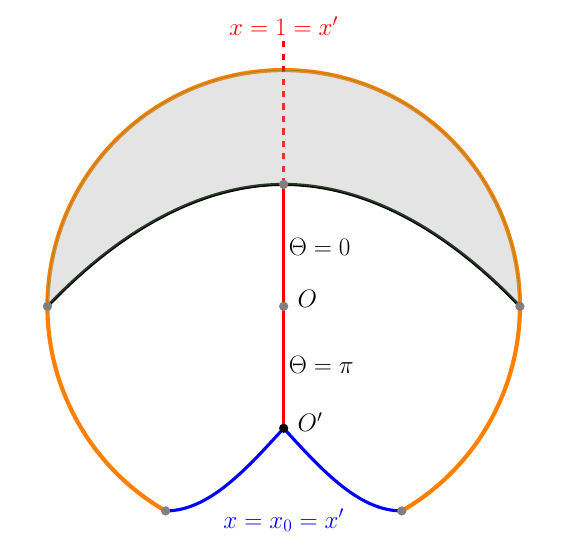}
  \caption{Engineer a conical geometry by gluing two EOW branes of the same color where we have mapped the C-metric to the \Poincare disk. By removing a wedge and identifying its edges $x=x_0=x'$, this resulting geometry corresponds to a slowly accelerating conical deficit sitting at $O'$ with $0<a<1$, while $O$ represents the AdS center and the black line corresponds to $r=\infty$. }\label{fig:gluing two copies of geometry}
\end{figure}

To describe a accelerating particle let us  consider the metric
\be 
ds^2=\frac{1}{(1+a r\cos(m\phi))^2}\left[-f(r)dt^2+\frac{dr^2}{f(r)}+m^2 r^2d\phi^2\right],
\ee
with $m<1,\, \phi\in[-\pi,\pi]$. The geometry has a conical singularity at $r=0$ with a deficit angle $2\pi(1-m)$ so it describes a static particle with mass $M=\frac{1-m}{4G_N}$ sitting at origin $r=0$.  Such a conical geometry can be engineered by gluing two non-maximal Class I solution as shown in Fig.\eqref{fig:gluing two copies of geometry}. The EOW brane at $x_0$ has a non-trivial tension and a corresponding non-vanishing on-shell action which exactly cancels \eqref{eq:onshelleow}. 
All the analysis about this geometry is parallel to one in the previous case so we leave the details in Appendix \eqref{appendix:conical}. 

\subsection{Class I in rapid phase: accelerating black hole}
The Class I C-metric can also describe a black hole geometry. Let us consider the Class I C-metric in the rapid phase where $a>1$, which has two Killing horizons at
\be 
y=\pm y_h,\quad y_h=\frac{\sqrt{a^2-1}}{a}.
\ee 
Except from the ranges of the parameters, the metric is same as the one in the slow phase, so the FG gauge and the holographic mass also have the same expressions.  We are particular interested in the question which patch may have a well-defined boundary dual?
For the simplest case with a trivial conformal structure, the holographic mass is proportional to the integral
\be 
M[\omega=w]\propto \int \frac{d\xi}{\sqrt{1-\xi^2}\Gamma}
\ee 
whose integrand has two singularities at $\xi=\pm y_h.$ Moreover,  for the transformation to FG gauge to be well-defined, the range of $\xi$ is $\xi<-y_h$ or $\xi>y_h$. Therefore, it is better to restrict ourselves into the range $y_h<x_0\leq \xi \leq 1$. Similarly, gluing two copies of such wedges results a black hole geometry with periodic spatial direction. The conformal anomaly only affects the holographic mass but not the temperature or the black hole entropy. The thermodynamics of this black hole with a general conformal factor can be studied with the method introduced in \cite{Tian:2023}. Here for simplicity we only consider the trivial conformal factor. The corresponding geometry is actually a saddle point solution.
 Extremizing the holographic mass (or the free energy) with respect to the conformal factor leads to the saddle point equation
\bea 
(1+3a^2(\xi^2-1))\xi \w \w' +(1-\xi^2)\Gamma(\xi )(\w'^2-\w \w'')=0,
\eea 
whose general solution is
\be 
\w=c_2 \exp\(c_1\frac{\arccoth\frac{\xi }{\sqrt{a^2-1}\sqrt{1-\xi^2}}}{\sqrt{a^2-1}}\).
\ee 
The trivial conformal factor solution corresponds to the case with $c_1=0$ and resulting the holographic mass \footnote{the factor comes from the counting of two copies of the wedge}
\bea  \label{eq:massrapid}
M=-\frac{1-a^2}{8\pi G_N}L,\quad L=\int_{x_0}^{1}\frac{d\xi }{\sqrt{1-\xi^2}\Gamma}=\frac{1}{\sqrt{a^2-1}}\arctanh \(\frac{\sqrt{a^2-1}\sqrt{1-x_0^2}}{x_0}\).
\eea  
From the surface gravity or the smooth condition of the Euclidean geometry, one can find that the temperature of the black hole is
\be \label{eq:class1tem}
T=\frac{\sqrt{a^2-1}}{2\pi}.
\ee 
As pointed out in \cite{Tian:2023}, a part of the black hole  Bekenstein-Hawking entropy which is proportional to the area of the horizon:
\bea 
S_{BH}&=&\frac{2}{4G_Na}\int_{x_0}^{1}\frac{dx}{(x-y_h)(\sqrt{1-x^2})},\\
&=&\frac{1}{2G_N}\arctanh[\frac{\sqrt{1-x_0^2}}{a-\sqrt{a^2-1}x_0}], \label{eq:class1BHentropy}
\eea  
is a boundary entropy coming from the EOW brane: 
\be 
S_{bdy}=\frac{c\times 2}{12}\log\frac{1+\mu}{1-\mu}=\frac{1}{2G_N}\log\sqrt{\frac{1+a\sqrt{1-x_0^2}}{1-a\sqrt{1-x_0^2}}},\quad \mu=a\sqrt{1-x_0^2}.
\ee 
It turns out that by subtracting out the boundary entropy we can formulate the Smarr relation:
\be \label{eq:class1smarr}
2M=T (S_{BH}-S_{bdy}),
\ee
where we have used the identity
\be 
\arctanh[\frac{\sqrt{1-x_0^2}}{a-\sqrt{a^2-1}x_0}]-\arctanh [\frac{\sqrt{a^2-1}\sqrt{1-x_0^2}}{x_0}]=\frac{1}{2}\log\frac{1+a\sqrt{1-x_0^2}}{1-a\sqrt{1-x_0^2}}.
\ee 
Varying the mass $M$ and the entropy $(S_{BH}-S_{bdy})$ with respect to $a$ and $L$ we can derive the first law of the thermodynamics:
\be\label{eq:firstlaw}
\delta M=T \delta (S_{BH}-S_{bdy})-\rho\delta L,
\ee
where $\rho=M/L$ can be interpreted as the energy density or a pressure. For the black hole with a generic conformal factor, we can add another term $\oint d\varphi_0\Omega_\varphi \delta \varphi$ in the first law to compensate the change of the conformal representative as we showed in \cite{Tian:2023}.
 
In the Appendix \eqref{appendix:rapid}, we will show the Euclidean on-shell action correctly reproduces the free energy.

\section{Holographic entanglement entropy}
In this section, we will use the RT formula to compute the holographic entanglement entropy of a single interval in the Class I geometry with particular emphasis placed on understanding how the acceleration, the EOW brane and the conformal factor affect the outcome. We will focus on the geometry described by a Class I wedge which is bounded by the EOW branes located $x=x_0$ and $x=1$ in the slow phase. According to the standard AdS/BCFT, this wedge is dual to a BCFT and the entanglement entropy of a single interval has two possible phases: the continuous phase and the discontinuous phase. In particular, the entanglement entropy in the continuous phase which corresponds to a continuous RT surface coincides with the one in a CFT which is supposed to be dual to the glued geometry.  In the discontinuous phase, the RT surface connects the interval and the EOW brane. In global AdS, there is an interesting phase transition between these two phases as we reviewed in the Appendix \eqref{appendix:phase}. We will explore a similar phase transition in the C-metric. By the way, We also noticed there exists a related calculation which is associated to regulated entropy and island prescription in 4 dimensional \cite{Landgren:2024ccz}.

\subsection{Continuous phase}
\begin{figure}[ht]
\centering 
  \includegraphics[scale=1.5]{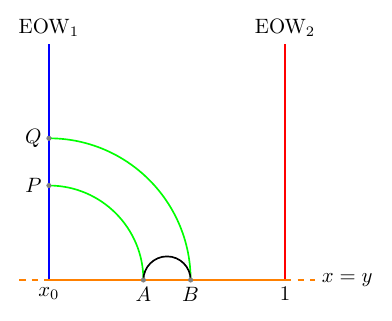}
  \caption{Connected and Disconnected geodesics given by AdS/BCFT duality. The black line which directly connects two end-points, and green lines which terminate on the EOW brane represent connected and disconnected geodesics respectively.}\label{fig:geodesic}
\end{figure}
Firstly let us consider the continuous phase. Let the two end points of the interval on the asymptotic boundary to be $A(t=0,y_1,x_1)$ and $B(t=0,y_2,x_2)$ and for convenience we will assume $x_2>x_1$. To be consist with the cut-off choice in the calculation of the on-shell action, we parameterize the asymptotic boundary as \eqref{eq:cutoff}
\be 
x-y=\Lambda(x)\epsilon=\frac{\sqrt{\Gamma(x)}}{\omega(x)}\epsilon.
\ee 
The continuous RT surface is just the geodesic connecting the two points. The simplest way to compute the geodesic length is to do the computation in the global coordinates \eqref{eq:global} in which the length of a geodesic $\Gamma$ anchored near the boundary at points $(R,T,\Theta)=(R_1,T_1,\Theta_1)$ and $(R,T,\Theta)=(R_2,T_2,\Theta_2)$ is given by
\be 
d_\Gamma=\log\(2R_1R_2[\cos(T_1-T_2)-\cos(\Theta_1-\Theta_2)]\).
\ee 
In the asymptotic limit, the transformation \eqref{eq:globalToC} implies that the coordinates the two points $A$ and $B$ in the global coordinates are
\bea 
T_i=0,\quad R_i=\frac{\omega(x_i)}{a\epsilon \sqrt{1-a^2}},\quad \cos\Theta_i=\frac{x_i}{\sqrt{\Gamma(x_i)}},\quad i=1,2,
\eea 
so the holographic entanglement entropy is 
\bea 
S_A&=&\frac{d_\Gamma}{4G_N}=\frac{c}{6}\log\left(\frac{2\omega(x_1)\omega(x_2)}{(a\epsilon)^2(1-a^2)}\right)\nonumber  \\
&+&\frac{c}{6}\log\(1-\frac{x_1x_2+(1-a^2)\sqrt{(1-x_1^2)(1-x_2^2)}}{\sqrt{\Gamma(x_1)\Gamma(x_2)}}\),
\eea 
which takes a rather simpler and more suggestive form in the conformal flat coordinates
\be 
S_{A,con}=\frac{c}{6}\log\left(\frac{\omega(X_1)\omega(X_2)}{(a\epsilon)^2(1-a^2)}\right)+\frac{c}{3}\log\(2\sin \frac{\sqrt{1-a^2}(X_2-X_1)}{2}\).
\ee 
A simple way to see this result is by observing the global AdS has the following asymptotic boundary metric  
\be 
ds^2=R^2(1-a^2)(-dt^2+\frac{d\Theta^2}{1-a^2})
\ee 
and comparing with \eqref{eq:standardFG}. The dependence of the conformal factor is exactly desired by a two-dimensional conformal field theory defined on a curved background \cite{Kikuchi:2019epb}. 
In the zero acceleration limit, the holographic entanglement entropy reduces to the one in the global AdS:
\bea 
S_{A,c}=\frac{c}{3}\log\frac{2\sin\frac{\f_1-\f_2}{2}}{\tilde{\e}},\quad \tilde{\e}=\left.\frac{\e}{\sqrt{\w(x_1)\w(x_2)}}\right|_{a \rightarrow 0}
\eea 
and in the limit $a\rightarrow 1$, the entropy reduces to the one in the \Poincare AdS:
\bea 
S_{A,c}=\frac{c}{3}\log\frac{\tan\f_1-\tan\f_2}{\tilde{\e}},\quad \tilde{\e}=\left.\frac{\e}{\sqrt{\w(x_1)\w(x_2)}}\right|_{a \rightarrow 1}
\eea 
where $x_i=\cos\phi_i$.

\subsection{Discontinuous phase}
In the discontinuous phase , the RT surface is piecewise as shown in Fig.\eqref{fig:geodesic}. The positions of the intersection should be determined by  minimizing individual geodesic segment. Taking the segment $AP$ for example, its length is 
\bea
d_{\Gamma_{AP}}&=&d_{\Gamma_{AP}} \\
&\approx &\log2\left(\sqrt{1+R_0^2}R_1-R_0R_1\cos[\Theta_1-\Theta_0]\right)
\eea 
where $R_0,\Theta_0$ are related by
\be 
\Theta_0=\arcsin\frac{\lambda}{R_0}+\Theta_{*},\quad \lambda\equiv-\frac{\sqrt{1-x_0^2}}{S^2+x_0^2}=\frac{\mu}{\sqrt{1-\mu^2}},\quad \cos\Theta_*=\frac{x_0}{\sqrt{\Gamma(x_0)}}
\ee 
Minimizing the geodesic length with respect to $R_0$ we find the saddle point is
\be
R_0=\sqrt{\frac{\lambda^2+\cos^2(\Theta_*-\Theta_1)}{\sin^2(\Theta_*-\Theta_1)}}
\ee 
and the corresponding geodesic length is
\bea 
d_{\Gamma_{A P}}=\log2\(\frac{\sqrt{\Gamma(x_1)}}{a\epsilon \sqrt{1-a^2} \Lambda(x_1)}(\sqrt{1+\lambda^2}+\lambda)\sin(\Theta_*-\Theta_1)\).
\eea 
Combining a similar term for the segment $BQ$, the total entanglement entropy in the discontinuous phase is
\bea 
S_{A,dis}&=&\frac{c}{3}\sinh ^{-1}(\lambda )+\frac{c}{6}\log\(\frac{\omega(x_1)\omega(x_2)}{(a\epsilon)^2(1-a^2)}\)+\frac{c}{6}\log\frac{2\sqrt{1-a^2}(x_1\sqrt{1-x_0^2}-x_0\sqrt{1-x_1^2})}{\sqrt{\Gamma(x_1)\Gamma(x_0)}}\nonumber \\
&+&\frac{c}{6}\log\frac{2\sqrt{1-a^2}(x_2\sqrt{1-x_0^2}-x_0\sqrt{1-x_2^2})}{\sqrt{\Gamma(x_2)\Gamma(x_0)}}.
\eea 
To compare it with the entropy in the continuous phase, we also rewrite it in terms of the conformal flat coordinates:
\bea 
S_{A,dis}-S_{A,con}=\frac{c}{3}\sinh ^{-1}(\lambda )+\frac{c}{6}\log\frac{\sin(\sqrt{1-a^2}\Delta)\sin(\sqrt{1-a^2}(\Delta+L))}{\sin(\sqrt{1-a^2}\frac{L}{2})}
\eea 
where $\Delta=X_1-X_0$ and $L=X_2-X_1$. Comparing with the results \eqref{eq:globald} in the global AdS, the only difference is that the lengths are rescaled by the factor $\sqrt{1-a^2}$ so the phase diagram is also similar to the one shown in Fig.\eqref{phase1}.

\section{The phase diagram}
In this section, we study the phase diagram of the three-dimensional C-metrics with a compact spatial dimension including the non-black-hole solutions: global AdS, I$_{\text{slow}}$ and the black hole solutions: BTZ \cite{Banados:1992wn}, Class I$_{\text{rapid}}$, Class II and Class III. For the purpose of comparison, we fix the boundary manifold to be a flat torus. All the bulk solution can be uniformly described by the metric:
\bea \label{eq:metric_static}
ds^2&=&\frac{dz^2}{z^2}+\frac{1}{z^2}(- dt^2+d\varphi^2)+\frac{u d\varphi^2+udt^2}{2}+z^2\frac{u^2d\varphi^2-u^2dt^2}{16},
\eea 
where  $u$ is only function of $\varphi$. In this convention, the free energy of all the possible bulk solution will be parameterized by two independent parameters: the size of the spatial circle $L_\varphi=\oint d\varphi$ and the temperature $T$.
The global AdS solution corresponds to the choice $u=-(2\pi/L_\varphi)^2$. The free energy is 
\be 
\mathfrak{F}_{AdS}=M_{\text{holo}}=\frac{1}{16\pi G_N}\int d\varphi \;u=-\frac{1}{16 \pi G_N }\frac{(2\pi)^2}{L_\varphi}.
\ee 
The BTZ solution corresponds to the choice of $u=u_{BTZ}>0$. The holographic mass, temperature and the entropy are
\bea 
&&M_{\text{holo}}= \frac{L_{\varphi}u_{BTZ}}{16\pi G_N},\quad T=\frac{\sqrt{u_{BTZ}}}{2\pi},\quad S=\frac{L_{\varphi} \sqrt{u_{BTZ}}}{4G_N}.
\eea 
Therefore, the free energy is 
\be \label{fbtz}
\mathfrak{F}_{BTZ}=-\frac{L_{\varphi}u_{BTZ}}{16\pi G}=-M_{\text{holo}}=-\frac{L_{\varphi}}{16\pi G_N}(2\pi T)^2=-\frac{1}{16\pi G_N}\frac{(2\pi)^2}{L_\varphi}(L_\varphi T)^2.
\ee 
According to \eqref{eq:holomass}, the free energy of the class I$_\text{slow}$ solution is
\be 
\mathfrak{F}_{\text{I}_{\text{slow}}}=-\frac{1 }{16\pi G_N} \int d\xi \frac{1-a^2}{\sqrt{1-\xi^2}\Gamma}=-\frac{\sqrt{1-a^2}}{8G_N}=-\frac{1}{16 \pi G_N }\frac{(2\pi)^2}{L_\varphi},
\ee 
where in the last step we have used the identity
\be 
{L_\varphi}=\int d\xi \frac{1}{\sqrt{1-\xi^2}\Gamma}=\frac{2\pi}{\sqrt{1-a^2}}\,,
\ee 
to express $a$ in terms of $L_\varphi$. Since $0<a<1$, the length $L_\varphi$ in the class I solution has a lower bound $L_\varphi\geq 2\pi$.
For the class I$_\text{rapid}$ solution, using \eqref{eq:massrapid}, \eqref{eq:class1tem} and \eqref{eq:class1smarr} we can rewrite the free energy as
\bea 
\mathfrak{F}_{\text{I}_{\text{rapid}}}=M-TS_{BH}=-\frac{1}{16\pi G_N}\frac{(2\pi)^2}{L_\varphi}(L_\varphi T)^2+\frac{T\arctanh[\sqrt{\frac{4\pi^2 T^2+1}{(2\pi T)^2\coth^2[L_\varphi\pi T]+1}}]}{2 G_N}.
\eea 
Similarly, as we show in the Appendix \eqref{appendix:class2} and \eqref{appendix:class3}, the free energy of class II solutions and class III solutions are
\bea
\mathfrak{F}_{II_\mp}&=&-\frac{1}{16\pi G_N}\frac{(2\pi)^2}{L_\varphi}(L_\varphi T)^2\pm\frac{T\arctanh[\sqrt{\frac{4\pi^2 T^2-1}{(2\pi T)^2\coth^2[L_\varphi\pi T]-1}}]}{2 G_N},\\
 \mathfrak{F}_{\text{III}}&=&-\frac{1}{16\pi G_N}\frac{(2\pi)^2}{L_\varphi}(L_\varphi T)^2.
\eea 
Here are some comments about the Class II solution. There are two types of Class II solution because a black hole geometry can be constructed by gluing two copies of Class II C-metric either with positive EOW brane tension or negative EOW brane tension. For a Class II black hole, the temperature is
\be 
T=\frac{\sqrt{a^2+1}}{2\pi}
\ee 
which has a minimal value $1/(2\pi)$ below which the Class II black hole does not exist. 
Now we are ready to compare the free energy of the these solutions. Among all the black hole solutions, it is easy to see that
\be 
\mathfrak{F}_{\text{II}_+}<\mathfrak{F}_{BTZ}=\mathfrak{F}_{\text{III}}<\mathfrak{F}_{\text{II}_-},\mathfrak{F}_{\text{I}_{\text{rapid}}}.
\ee 

The two non-black-hole solutions have the same free energy but the Class I$_{\text{slow}}$ solution only exists when $L_\varphi\geq 2\pi$. This is consistent with our expectation that the Class I$_{\text{slow}}$ solution interpolates between the global AdS and \Poincare AdS when we tune the parameter $a$.

\begin{figure}[ht]
\centering 
  \includegraphics[scale=0.3]{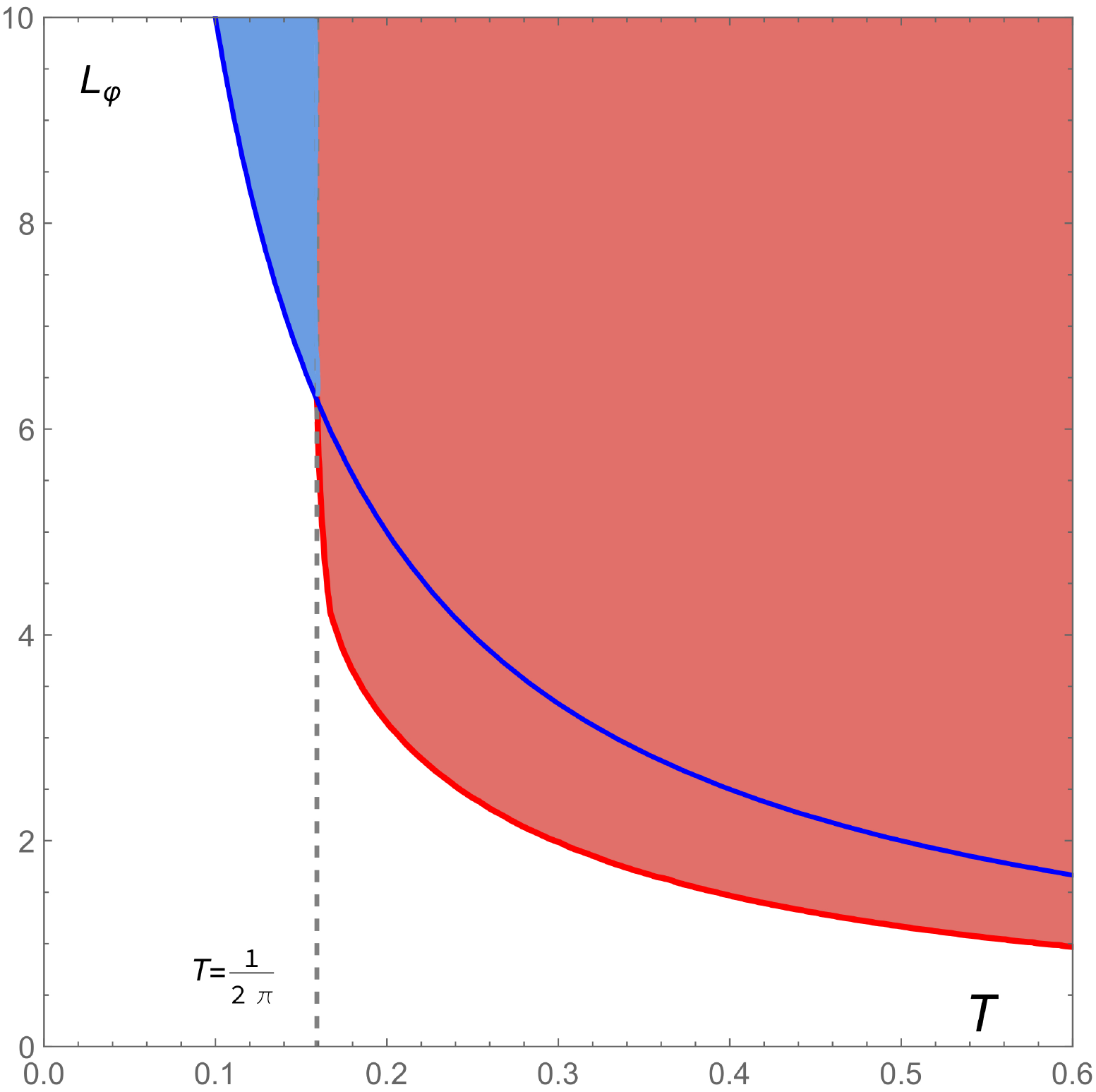}
  \caption{The phase diagram. The blue curve denotes the standard Hawking-Page temperature. The red curve represents the solution of the \eqref{eq:transition}. The region in \blue{blue} is the BTZ phase and the region in \red{red} is the Class II$_+$ accelerating black hole phase.}\label{fig:phase}
\end{figure}

When the temperature is below $1/(2\pi)$, Class II$_{+}$ solution does not exist so the phase transition takes place at the Hawking-Page temperature $T=T_{HP}=\frac{1}{L_\varphi}$. However, when the temperature is above $1/(2\pi)$ the Class II$_+$ exhibits a lower free energy due to its larger Bekenstein-Hawking entropy, which includes an additional boundary entropy arising from the EOW brane construction. The phase transition temperature should be determined by the transcendental equation
\be\label{eq:transition}
(L_\varphi T)^2\pi+2(L_\varphi T)\arctanh[\sqrt{\frac{4\pi^2 T^2-1}{(2\pi T)^2\coth^2[L_\varphi T\pi ]-1}}]=\pi\, .
\ee 
We plot the phase diagram in Fig.\eqref{fig:phase}. When $L_\varphi>2\pi$, we will encounter three phases: the global AdS(or Class I$_{\text{slow}}$), the BTZ black hole and the Class II$_+$ accelerating black hole consecutively as the temperature increases. There exists a triple critical point at $(L_\varphi=2\pi,\, T=1/(2\pi))$. Above the critical temperature $T=1/(2\pi))$, the phase transition temperature is always lower than the Hawking-Page temperature.

\section{Conclusions and Outlooks}
In this study, we systematically investigated the thermodynamic and holographic properties of 3D C-metrics in the FG gauge. Despite the ambiguity in choosing the conformal representative, we demonstrated that the free energy equals the Euclidean on-shell action, and the entanglement entropy behavior aligns with field theory analysis expectations. By factoring out the conformal factor dependence in the holographic mass and identifying the boundary entropy contribution to the black hole Bekenstein-Hawking entropy, we successfully formulated the first law of thermodynamics and the Smarr relation. We also depicted the phase diagram of bulk solutions with a fixed flat torus boundary geometry. The accelerating black hole dominates the BTZ black hole when the temperature exceeds a critical value. We anticipate similar analyses for other 3D accelerating black holes, such as the charged \cite{EslamPanah:2022ihg,Jafarzade:2017kin} and scalar hair varieties \cite{Cisterna:2023qhh} . Some open questions remain.

While the presence of boundary entropy is natural due to the EOW brane construction, its holographic interpretation in this context remains unclear. We anticipate its appearance in other glued geometries, such as those in \cite{Kawamoto:2023wzj}. Boundary entropy typically relates to a BCFT or can be viewed as part of the regularized \Renyi entropy. As boundary entropy can be negative, it is possible for the total Bekenstein-Hawking entropy to become negative. A similar phenomenon has been observed for the \Renyi entropy, linked to bulk solutions with intersecting EOW branes \cite{Tian:2023vbi}.

Further exploration of the consequences of the conformal factor ambiguity is necessary. For a generic choice of the conformal factor, the boundary metric is curved. Recent progress on 2D CFT defined on curved spacetime is intriguing \cite{deBoer:2023lrd}. Presumably, C-metrics can serve as a concrete toy holographic model. 

The relationship between boundary conditions and the conformal representative can be derived in the Chern-Simons formalism of AdS gravity, as shown in \cite{Tian:2023}, following \cite{Perez:2016vqo,Ojeda:2019xih,Dymarsky:2020tjh}. In this formalism, we can define a thermodynamic mass different from the holographic mass studied in this work, leading to a different system thermodynamics. We leave this intriguing problem for future investigation.

\section*{Acknowledgments}
We thank Huajia Wang, Cheng Peng, and Nozaki Masahiro for the valuable discussion and collaboration on related topics. JT is supported by  the National Youth Fund No.12105289.

\appendix
\section{The cut-off}
\label{appendix:cut-off}
According to FG expansion, at leading order
\bal
x=\x+F_{1}(\x)\,z,\quad y=\x+G_{1}(\x)\,z
\eal
with
\bal
G_{1}=-\frac{a^2(1-\x^2)}{\G}F_{1}, \quad \G=1-a^2\(1-\x^2\)
\eal
in slow phase, which means
\bal
F_{1}-G_{1}=\frac{F_{1}}{\G}
\eal
so it's direct to solve $\x$ at leading order 
\bal
\x=\x_1=\frac{-1\pm\sqrt{1+4a^2(x-y)\(a^2(x-y)+y\)}}{2a^2(x-y)}
\eal
here the label $1$ represents this is a solution at leading order which is only reasonable at leading order. Since we expect $\x$ is identical to $x$ and $y$ at leading order if we expand the metric near the boundary and in order to cover the $x\geq 0$ region, so we will take the solution with plus sign. Substitute this solution into the above cutoff and take $z=\e$
\bal
x-y=\frac{F_{1}(\x_1)}{\G(\x_1)}\,\e
\eal
since 
\bal
\x_1=x-\G(x)\(x-y\)+\mathcal{O}(\(x-y\)^2)
\eal
, a naive and reasonable cutoff at leading order should be taken as follows
\bal\label{FG cutoff}
x-y=\L(x)\e,\quad \L(x)=\frac{F_1(x)}{\G(x)}
\eal

\section{On-shell action of the conical geometry}
\label{appendix:conical}
In this appendix, we compute the on-shell action of the conical geometry which is constructed by gluing two non-maximal Class I C-metrics in the slow phase.

The bulk Einstein-Hilbert action is
	\bal
	I_N&=-\frac{1}{16\pi G_N}\int d^3x \sqrt{g}\,\(R+2\)
	\eal
where the determinant of the metric is
	\bal
	 \sqrt{g}=\frac{1}{a^2\(x-y\)^3\sqrt{1-x^2}}
	\eal
 so the above integral is evaluated as 
 \bal
 I_{N}&=\frac{1}{4\pi G_N a^2}\int dt_E \int_{x_0}^{1}dx\frac{2\times 1}{\sqrt{1-x^2}}\int_{-\infty}^{x-\frac{\L(x)\e}{a}}\frac{dy}{\(x-y\)^3}\nt
 &=\frac{1}{4\pi G_N a^2}\int dt_E \int_{x_0}^{1}\frac{dx}{\sqrt{1-x^2}}\frac{a^2}{{\L(x)}^2\e^2}
 \eal
the factor $2$ takes account of two copies of the geometry. The Brown-York action with the counter term is
	\bal
	I_{bdy}=-\frac{1}{8\pi G_N}\int d^2x\,\sqrt{h}\(K-1\),
	\eal
where the integrand is
\bal
\sqrt{h}\(K-1\)=\frac{1}{a^2\e^2}\,\frac{a^2}{{\L(x)}^2\sqrt{1-x^2}}+\frac{1}{\e}\frac{d\(\frac{a\sqrt{1-x^2}}{\L(x)}\)}{dx}+U(x),
\eal
and $U(x)$ is a very complicated function
\bal \label{eq:ux}
U(x)=\frac{1-a^2}{2}\frac{1}{\G\sqrt{1-x^2}}+\frac{3\G\sqrt{1-x^2}\,{\w'}^2}{2\,\w^2}+\frac{x\G\(3\G-2\)}{\,\G\sqrt{1-x^2}}\frac{\w'}{\w}-\frac{\G\sqrt{1-x^2}\w''}{\w}.
\eal	
As we mentioned in the section \ref{sec:on_shell}, the new sub-leading $\epsilon^{-1}$ divergence will be canceled by the EOW brane action:
	\bal
	I_{Q}&=-\frac{2\m_0}{8\pi G_N}\int d^2x \sqrt{h}\nt
	&=-\frac{\m_0}{4\pi G_N}\int dt_E\int_{-\infty}^{x_0-\frac{\L(x_0)\e}{a}}\frac{dy}{a(x_0-y)^2}\nt
	&=-\frac{\mu_0}{4\pi G_N}\int dt_E\frac{1}{\L(x_0)\e} \label{eq:eowaction }
	\eal
where the brane tension is $\m_0=a\sqrt{1-x_0^2}$. For convenience we would like to introduce an extra term which has no contribution to the on-shell action but will simplify the final result a lot
\bal
I_{extra}&=-\frac{2\times 1}{16\pi G_N}\int dt_E \int_{-1}^{1}\,\frac{d}{dx}\[-\frac{2\sqrt{1-x^2}\,\G \w'}{\w}\]dx\nt
&=\frac{1}{4\pi G_N}\int dt_E \[\int_{x_0}^{1}\,\frac{d}{dx}\(\frac{\sqrt{1-x^2}\,\G \w'}{\w}\)dx+\int_{-1}^{x_0}\,\frac{d}{dx}\(\frac{\sqrt{1-x^2}\,\G \w'}{\w}\)dx\]\label{eq:totalextra}
\eal  
Summing over all these terms leads to  the final result 
\bal
I&=I_{N}+I_{Q}+I_{ct}-I_{extra}\nt
&=-\frac{1}{4\pi G_N}\int dt_E \[\int_{x_0}^{1}\frac{1}{2}\(\frac{1-a^2}{\sqrt{1-x^2}\,\G}+\frac{\G\sqrt{1-x^2}\,{\w'}^2}{\w^2}\)dx+\left.\(\frac{\sqrt{1-x^2}\,\G \w'}{\w}\)\right|_{x=x_0}\],\\
&=\int dt_E M[\w],
\eal
recalling that the holographic mass is
	\bal
M[\w]&=-2\int_{x_0}^{1}d\x\sqrt{g^{(0)}}T^{t}_{\;t}\nt	
&=-\frac{1}{4\pi G_N}\int_{x_0}^{1} d\x\(\frac{1-a^2}{2}\frac{1}{\G\sqrt{1-\x^2}}+\frac{1}{2}\,\frac{\G\sqrt{1-\x^2}\,{\w'}^2}{\w^2}\)-\frac{1}{4\pi G_N}\left.\(\frac{\G\sqrt{1-\x^2}\w'}{\,\w}\)\right|_{\x=x_0}.
\eal

\section{On-shell action of the Class I C-metric in rapid phase}
\label{appendix:rapid} 
In this appendix we compute the on-shell action of Class I C-metric in the rapid phase. Some of the steps and results are similar to the ones in Appendix \ref{appendix:conical}.
The Einstein-Hilbert on-shell action is
 \bal
 I_{N}&=\frac{1}{4\pi G_N}\int d^3 x\sqrt{g}\nt
 &=\frac{1}{4\pi G_N}\frac{1}{a^2}\int dt_E \int_{x_0}^{1}\frac{2\times 1}{\sqrt{1-x^2}}dx\int_{y_h}^{x-\frac{\L(x)\e}{a}}\frac{dy}{(x-y)^3}\nt
 &=\frac{1}{4\pi G_N}\frac{1}{a^2}\int dt_E \int_{x_0}^{1}\frac{1}{\sqrt{1-x^2}}\(\frac{a^2}{\L(x)^2\e^2}-\frac{1}{\(x-y_h\)^2}\)dx.
 \eal
The second term can be evaluated as
 \bal
 \int _{x_0}^{1}\frac{dx}{\sqrt{1-x^2}\(x-y_h\)^2}=\frac{\sqrt{1-x_0^2}}{(x_0-y_h)\(1-y_h^2\)}+\frac{2y_h\arctanh\sqrt{\frac{(1+y_h)(1-x_0)}{(1-y_h)(1+x_0)}}}{\(1-y_h^2\)^{\frac{3}{2}}}
 \eal 
The Brown-York action with the counter term is
\bal
I_{ct}&=-\frac{1}{8\pi G_N}\int d^{2}x\sqrt{h}\(K-1\)\nt
&=-\frac{2\times 1}{8\pi G_N}\int dt_E\int_{x_0}^{1} dx \[\frac{1}{a^2\e^2}\,\frac{a^2}{{\L(x)}^2\sqrt{1-x^2}}+\frac{1}{\e}\frac{d\(\frac{\sqrt{1-x^2}}{a\L(x)}\)}{dx}+U(x)\].
\eal 	 
here $U(x)$ is same as \eqref{eq:ux}. To cancel the $\epsilon^{-1}$ divergence we also need to include the EOW brane action \eqref{eq:eowaction }.
Subtracting the total derivative term \eqref{eq:totalextra} for simplicity, the final result of on-shell action is
\bal
I&=I_{N}+I_{Q}+I_{ct}-I_{extra}\nt
&=-\frac{1}{4\pi G_N}\int dt_E \int_{x_0}^{1}\frac{1}{2}\(\frac{1-a^2}{\sqrt{1-x^2}\,\G}+\frac{\G\sqrt{1-x^2}\,{\w'}^2}{\w^2}\)dx - \frac{1}{4\pi G_N }\int dt_E \left.\frac{\sqrt{1-x^2}\,\G \w'}{\,\w}\right|_{x=x_0}\nt
&\phantom{\;\;}-\int dt_E \frac{ay_h \arctanh\sqrt{\frac{(1+y_h)(1-x_0)}{(1-y_h)(1+x_0)}}}{2\pi G_N}.
\eal
Recalling \eqref{eq:holomass},\eqref{eq:class1BHentropy} and \eqref{eq:class1tem}, we immediately obtain
\bal\label{statistical relation}
I=\b\(M- T S_{BH}\).
\eal

\section{Continuous phase vs. discontinuous phase}
\label{appendix:phase}
\begin{figure}[h]
\centering
  \includegraphics[scale=0.4]{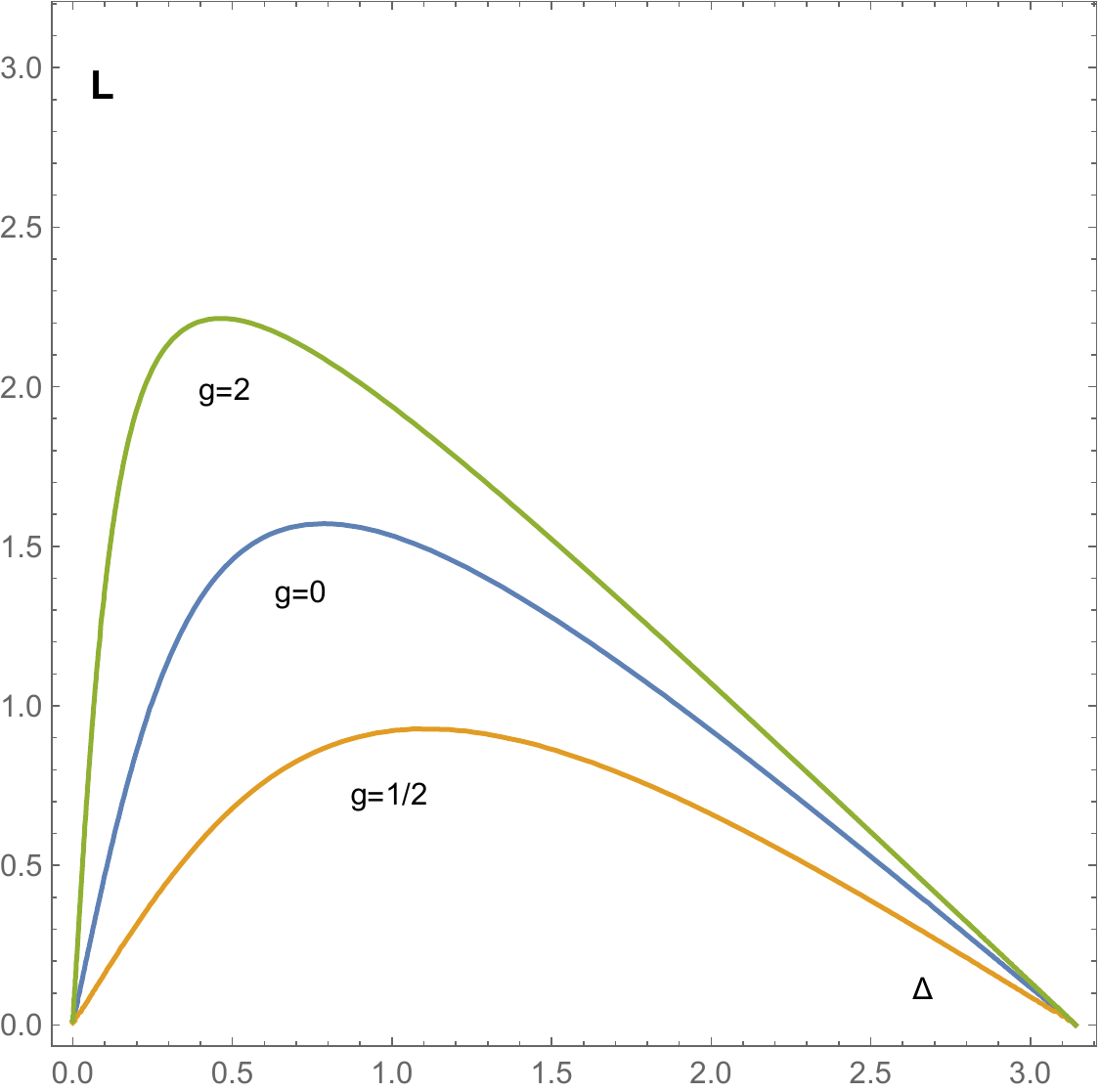}
  \caption{The region below the contour is where $d_{dis}>d_{con}$ and different color corresponds to the case with different boundary entropy.}\label{phase1}
\end{figure}

In this section, we study continuous/discontinuous phase transition of a short interval in the global AdS space with a EOW brane. Let the two end points of the interval and the EOW brane end point on the boundary to be $\Theta_2>\Theta_1$ and $\Theta_0$, respectively. Without the loss of generality, we can assume $\Theta_2>\Theta_1>\Theta_0$. As geodesic distance is isometry-invariant, it can be read directly via embedding coordinate
\bal
\cosh d=-X\cdot X'
\eal
substitute the transformation between ambient space and global AdS$_3$, the continuous geodesic length is
\bea 
d_{con}=2\log\(2\sin\frac{\Theta_2-\Theta_1}{2}\)-2\log \epsilon,
\eea 
while the discontinuous geodesic length is
\bea 
d_{dis}=2\arcsinh(\lambda )+2\log [2\sqrt{\sin(\Theta_2-\Theta_0)\sin(\Theta_1-\Theta_0)}]-2\log\epsilon
\eea 
Let us parameterize the geodesic in terms of the length of the interval $L=\Theta_2-\Theta_1$ and the distance between the interval and the EOW brane $\Delta=\Theta_1-\Theta_0$. Then the difference of the two geodesics is
\bea \label{eq:globald}
d_{dis}-d_{con}=2\log\frac{\sqrt{\sin\Delta \sin(\Delta+L)}}{\sin\frac{L}{2}}+2\log g,\quad g=\lambda+\sqrt{1+\lambda^2}
\eea 
The phase diagram is shown in \eqref{phase1}.

\section{Class II solution}
\label{appendix:class2}
The metric of the Class II solution in the FG gauge is
\be
g^{(0)}_{ij}dx^idx^j=-\omega(x)^2dt^2+\frac{\omega(x)^2dx^2}{(x^2-1)\Gamma^2},\quad \Gamma=1+a^2-a^2x^2,
\ee 
and
\bea 
&&g_{tt}^{(2)}=\frac{1+a^2}{2}-\frac{(x^2-1)\Gamma^2\omega'^2}{2\omega^2},\\
&&g_{x x}^{(2)}=\frac{1+a^2}{2(x^2-1)\Gamma^2}+\frac{x (1-3a^2(x^2-1))}{(x^2-1)\Gamma}\frac{\omega'}{\omega}+\frac{2\omega\omega''-3\omega'^2}{2\omega^2},\\
&&g^{(4)}=\frac{1}{4}g^{(2)}{g^{(0)}}^{-1}g^{(2)}.
\eea 
The metric of the asymptotic boundary of Class II C-metric is 
\be 
ds^2_{bdy}={\omega(Y)^2}{}\(-dt^2+dY^2\),\quad dY=\frac{dx}{\sqrt{x^2-1}(1+a^2-a^2 x^2)}.
\ee 
\subsection{Class II$_-$}

The Class II$_-$ black hole is constructed by gluing two copies of the wedge bounded by the EOW brane $x=x_0>0, x=1$, the horizon $y=-y_h=-\sqrt{1+a^2}/a$ and the conformal boundary $x=y$. The holographic mass is
\be 
M_{\text{holo}}[\omega]=\frac{1}{16\pi G_N} \int dY\, \(a^2+1+\frac{2\Omega\Omega''-3\Omega'^2}{\Omega^2} \).
\ee 
When the conformal factor is trivial, the mass reduces to
\be \label{eq:2pm_1}
M_{\text{holo}}[1]=\frac{\sqrt{1+a^2}}{8\pi G_N}{\arctanh (\frac{\sqrt{(1+a^2)(x_0^2-1)}}{x_0})}{}.
\ee 
The temperature of the black hole is
\be 
T=\frac{\sqrt{1+a^2}}{2\pi}.
\ee 
The Bekenstein-Hawking entropy is 
\be 
S_{BH}=\frac{2}{4G_Na}\int^{x_0}_{1}\frac{dx}{(x+y_h)(\sqrt{x^2-1})}=\frac{1}{2G_N}\arctanh\left[\frac{\sqrt{x_0^2-1}}{a+\sqrt{a^2+1}x_0}\right]
\ee 
and the boundary entropy is
\be 
S_{bdy,-}=\frac{c\times 2}{12}\log\frac{1+\mu}{1-\mu}=\frac{1}{2G_N}\log\sqrt{\frac{1-a\sqrt{x_0^2-1}}{1+a\sqrt{x_0^2-1}}},\quad \mu=-a\sqrt{x_0^2-1}.
\ee 
So the black hole satisfies the Smarr relation
\be 
2M_{\text{holo}}[1]=T(S_{BH}-S_{bdy}).
\ee 
Thus, the free energy is given by
\be 
\mathfrak{F}_{\text{Class II$-$}}=M_{\text{holo}}[\Omega]-(2M_{\text{holo}}[1]+TS_{bdy,-})
\ee 
When the conformal factor is trivial, the boundary size is
\be 
L_{\varphi}=\int dY=\frac{16\pi G_N M_{\text{holo}}[1]}{1+a^2}=\frac{2 \tanh ^{-1}\left(\frac{\sqrt{\left(a^2+1\right) \left(x_0^2-1\right)}}{x_0}\right)}{\sqrt{a^2+1}}\in[0,\infty).
\ee 
\subsection{Class II$_+$}
The Class II$_+$ black hole is constructed by gluing two copies of the wedge bounded by the EOW brane $x=-x_0<0, x=-1$, the horizon $y=-y_h$ and the conformal boundary $x=y$. The holographic mass is still given by
\be 
M_{\text{holo}}[\omega]=\frac{1}{16\pi G_N} \int dY\, \(a^2+1+\frac{2\Omega\Omega''-3\Omega'^2}{\Omega^2} \),
\ee 
and \eqref{eq:2pm_1}. The temperature is also the same but the Bekenstein-Hawking entropy becomes
\be 
S_{BH}=\frac{1}{2G_N}\arctanh\left[\frac{\sqrt{x_0^2-1}}{\sqrt{1+a^2}x_0-a}\right]
\ee 
and the boundary entropy is
\be 
S_{bdy,+}=\frac{c\times 2}{12}\log\frac{1+\mu}{1-\mu}=\frac{1}{2G_N}\log\sqrt{\frac{1+a\sqrt{x_0^2-1}}{1-a\sqrt{x_0^2-1}}},\quad \mu=a\sqrt{x_0^2-1}.
\ee 
The black hole satisfies the Smarr relation
\be 
2M_{\text{holo}}[1]=T(S_{BH}-S_{bdy,+}).
\ee 
Thus, the free energy is given by
\be 
\mathfrak{F}_{\text{Class II$+$}}=M_{\text{holo}}[\Omega]-(2M_{\text{holo}}[1]+TS_{bdy,+}).
\ee 
Note that $S_{bdy,+}>0>S_{bdy,-}$ so at the same temperature $\mathfrak{F}_{\text{Class II$+$}}<\mathfrak{F}_{\text{Class II$-$}}$.
When the conformal factor is trivial, the boundary size is
\be 
L_{\varphi}=\int dY=\frac{16\pi G_N M_{\text{holo}}[1]}{1+a^2}.
\ee 
\section{Class III solution}
\label{appendix:class3}
First, let us rewrite the metric in the FG gauge.
\bea
&&g_{ij}^{(0)}dx^idx^j=\frac{\omega^2}{a^2}\(-d\tau^2+\frac{a^2 d\xi^2}{(1+\xi^2)\Gamma^2}\),\\
&&\Gamma=1-a^2(1+\xi^2)
\eea 
\bea 
&&g_{\tau\tau}^{(2)}=\frac{1-a^2}{2a^2}-\frac{(1+\xi^2)\Gamma^2\omega'^2}{2a^2\omega^2},\\
&&g_{\xi\xi}^{(2)}=\frac{1-a^2}{2(1+\xi^2)\Gamma^2}+\frac{\xi(1-3a^2(1+\xi^2))}{(1+\xi^2)\Gamma}\frac{\omega'}{\omega}+\frac{2\omega\omega''-3\omega'^2}{2\omega^2}
\eea 
The holographic mass is given by
\be 
M=\frac{1}{16\pi G_N}\int d\xi \(\frac{1-a^2}{\sqrt{1+\xi^2}\Gamma}-\frac{\sqrt{1+\xi^2}\Gamma\omega'^2}{\omega^2}\).
\ee 
Note that for the transformation to the FG gauge to be real, we have to restrict $x$ into the region
\be 
x^2\leq \frac{1-a^2}{a^2},\quad 0<a<1.
\ee 
However, if we include the whole region the holographic mass will diverge. So let us put two EOW branes at $x=\pm x_0$. Setting $\omega=w$, the holographic mass then is
\bea
M=\frac{1}{16\pi G_N}\int_{-x_0}^{x_0}\(\frac{1-a^2}{\sqrt{1+\xi^2}\Gamma}\)=\frac{\sqrt{1-a^2}}{8\pi G_N}{\arctanh[\frac{x_0}{\sqrt{1-a^2}\sqrt{1+x_0^2}}]}{}.
\eea 
The temperature of the black hole is 
\be 
T=\frac{\sqrt{1-a^2}}{2\pi}.
\ee
The entropy is 
\bea 
S_{BH}&=&\frac{1}{4G_N a}\int_{-x_0}^{x_0}\frac{dx}{(x+y_h)\sqrt{1+x^2}},\\
&=&\frac{1}{4G_N}\left(\tanh ^{-1}\left(\frac{\sqrt{1-a^2} x_0-a}{\sqrt{x_0^2+1}}\right)+\tanh ^{-1}\left(\frac{\sqrt{1-a^2} x_0+a}{\sqrt{x_0^2+1}}\right)\right),\\
&=&\frac{1}{2G_N}\arctanh[\frac{x_0}{\sqrt{1-a^2}\sqrt{1+x_0^2}}].
\eea 
Therefore, we have the standard Smarr relation
\be 
2M=TS_{BH}.
\ee 
\printbibliography

\end{document}